**Major Solar Eruptions and High Energy Particle Events during Solar Cycle 24**

N. Gopalswamy[1]
NASA Goddard Space Flight Center, Greenbelt, Maryland, USA
H. Xie[2], S. Akiyama[3], P. Mäkelä[4], and S. Yashiro[5]
The Catholic University of America, Washington DC 20064, USA
[1]Corresponding author: nat.gopalswamy@nasa.gov
[2]hong.xie@nasa.gov
[3]Sachiko.Akiyama@nasa.gov
[4]pertti.a.makela@nasa.gov
[5]seiji.yashiro@nasa.gov

**Abstract**
We report on a study of all major solar eruptions that occurred on the frontside of the Sun during the rise to peak phase of cycle 24 (first 62 months) in order to understand the key factors affecting the occurrence of large solar energetic particle events (SEPs) and the ground levels enhancement (GLE) events. The eruptions involve major flares with soft X-ray peak flux $\geq 5.0 \times 10^{-5}$ Wm$^{-2}$ (i.e., flare size $\geq$M5.0) and accompanying coronal mass ejections (CMEs). The selection criterion was based on the fact that the only front-side GLE in cycle 24 (GLE 71) had a flare size of M5.1. Only ~37% of the major eruptions from the western hemisphere resulted in large SEP events. Almost the same number of large SEP events was produced in weaker eruptions (flare size <M5.0), suggesting that the soft X-ray flare is not a good indicator of SEP or GLE events. On the other hand, the CME speed is a better indicator of SEP and GLE events because it is consistently high supporting the shock acceleration mechanism for SEPs and GLEs. We found the CME speed, magnetic connectivity to Earth, and ambient conditions as the main factors that contribute to the lack of high energy particle events during cycle 24. Several eruptions poorly connected to Earth (eastern-hemisphere or behind-the-west-limb events) resulted in very large SEP events detected by the STEREO spacecraft. Some very fast CMEs, likely to have accelerated particles to GeV energies, did not result in a GLE event because of poor latitudinal connectivity. The stringent latitudinal requirement suggests that the highest energy particles are likely accelerated in the nose part of shocks. There were also well-connected fast CMEs, which did not seem to have accelerated high energy particles due to possible unfavorable ambient conditions (high Alfven speed, overall reduction in acceleration efficiency in cycle 24).

**1. Introduction**

The Ground Level Enhancement (GLE) in solar energetic particle (SEP) events was first detected in 1942 (Forbush 1946). Since then there have been 72 GLE events in the past 72 years, amounting to about a dozen events during each solar cycle. GLEs represent the highest energy particles in SEP events often exceeding ~1 GeV that may have important implications for Earth over various timescales (see e.g. Shea and Smart 2012; Lammer et al. 2012). Coronal mass ejections (CMEs), discovered in 1971 (Tousey, 1973), are now thought to be the source of large SEP and GLE events consisting of particles accelerated by the CME-driven shock (Kahler et al. 1978; Cliver 2006, Gopalswamy et al. 2012). Although flare reconnection process has also been thought to be a candidate for accelerating GLE particles (Bazilevskaya 2008; Grechnev et al. 2008), the association of some GLE events with weak flares makes this mechanism less viable. Therefore, the shock mechanism for GLEs has emerged to the forefront.

There were 16 GLE events during solar cycle 23 (1996 – 2008) all associated with the very high energy CMEs (Gopalswamy et al. 2012). However, there have been only two GLE events (GLE71 on 2012 May 17 and GLE72 on 2014 January 6) during the first 5 years of cycle 24 (Gopalswamy et al. 2013a, 2014; Thakur et al. 2014). GLE71 was associated with a M5.1 flare and a very fast (~2000 km/s) CME. GLE72 was from a source region behind the west limb, so we do not have flare information (Thakur et al. 2014). However, the CME was very fast (~1700 km/s). Thus the two GLE events of cycle 24 are consistent with shock acceleration. However, there were other cycle-24 energetic eruptions (larger flares and faster CMEs) from the same longitude range as GLE71 that did not result in GLEs (Gopalswamy et al. 2013a, paper 1). One possibility is the poor latitudinal connectivity to Earth for non-GLE events: the average latitudinal distance from the ecliptic was 32º for the cycle-24 non-GLE events compared to 13º for the cycle-23 GLE events. The two cycle-24 GLE sources were within 5º from the ecliptic, suggesting excellent latitudinal connectivity. Favorable solar B0 angle and/or non-radial CME motion seem to be two of the reasons for the occurrence of several historical GLE events with flare latitudes >30º (Gopalswamy and Mäkelä 2014), consistent with the above result. The B0 angle is the latitude of the ecliptic in heliographic coordinates (i.e., inclination of the solar equator to the ecliptic) and has values in the range ±7º.25, positive (negative) referring to the north (south).

Paper 1 was primarily on GLE71 and examined why large eruptions of similar flare magnitude (≥M5.0) and source longitudes (W55-W90) did not result in high energy particles. Since GLE events do originate from outside this longitude range (albeit with lower probability), we need to consider major eruptions from other longitudes. Most GLE sources are located to the west of E15, and only two (or <3%) are known to have occurred at large eastern longitudes: GLE #36 from E31 (October 12, 1981 – Cliver 2006) and GLE #9 from E88 (September 3, 1960 – Cliver et al. 1982). In this paper, we examine major flares (≥M5.0) and the associated CMEs, extending (i) the longitude range considered in Paper 1 and (ii) the study period to 2014 January 31. We also consider other eruptions (flare size <M5.0), that were associated with large SEP events; large SEP events are those with a proton intensity in >10 MeV GOES energy channel exceeding 10 pfu [particle flux units; 1 pfu = 1 particle per (cm$^2$ s sr)].

The ultimate aim is to find additional causes that might contribute to SEP variability and explain the rarity of GLE events. These include the overall reduction in the rate of energetic eruptions due to the weak solar activity in cycle 24, poor latitudinal connectivity, and the reduced efficiency of shock acceleration in the modified heliosphere. It was recently shown that the reduction in the number of energetic eruptions in cycle 24 (compared to cycle 23) may not account for the extremely low rate of GLE events (Gopalswamy et al. 2014).

## 2. Data Selection and Analysis

We considered all flares reported in on-line Solar Geophysical Data that have a soft X-ray (SXR) peak flux ≥ M5.0. The solar source locations of these flares are generally given as the location of the associated H-alpha flare by the Space Weather Prediction Center (SWPC). We consider only disk events, because the true size of limb and backside flares is unknown. For flares that do not have their solar source location listed, we examined movies of EUV solar images obtained by the Solar Dynamics Observatory's Atmospheric Imaging Assembly (AIA, Lemen et al. 2012) to identify the eruption location. We also made use of movies obtained by the Extreme Ultraviolet Imager (EUVI) on board the Solar Terrestrial Relations Observatory (STEREO) to check if a source is front-or back-sided. In all, 69 flares with SXR peak flux ≥M5.0 were reported by SWPC. Once we decided on the solar sources of all the flares, we looked for the associated CMEs in the coronal images obtained by the Large Angle and Spectrometric Coronagraph (LASCO, Brueckner et al. 1995) on board the Solar and Heliospheric Observatory (SOHO) spacecraft. We made use of the measurements available from the SOHO/LASCO CME catalog at the CDAW Data Center (http://cdaw.gsfc.nasa.gov, Gopalswamy et al. 2009a) for the associations. For CMEs that occurred during the last seven months (July 2013 to January, 2014) we made measurements using preliminary data. Of the 69 flares, 9 were not associated with a CME. These included 7 M- and 2 X-class flares. These confined flares involve no mass motion (Gopalswamy et al. 2009b) and hence are not relevant for this study. For one flare (2012 May 10 at 04:11 UT from N12E22), there were no SOHO/LASCO observations, but the CME was observed by STEREO-Behind (STB). We excluded this event also because of the incomplete data. The remaining 59 flares and the associated CMEs form the primary data set for this study.

We also used the set of all large SEP events of cycle 24, 31 in all, until the end of January 2014. We compiled the flare and CME information for these SEP events, extending an earlier report (Gopalswamy, 2012). Only 16 of the 31 large SEP events were associated with the major eruptions in the primary data set. The remaining 15 SEP events were either back-sided or associated with weaker flares (flare size < M5.0). The overlap between the two data sets is summarized in Table 1.

We subdivided the list of 59 major eruptions as follows: (i) 16 eruptions associated with large SEP events (Table 2), (ii) 24 eruptions from eastern longitudes (east of E15) (Table 3), and (iii) 22 non-SEP eruptions from GLE longitudes (E15 – W90) (Table 4). The subset (i) is the primary candidate list for GLE association because there is a weak correlation between SEP intensity and GLE intensity (Gopalswamy et al. 2012). We do not expect GLEs from subset (ii) because they are eastern events but may be associated with SEP events. Events in subset (iii) were from GLE longitudes, but they lacked SEP events. It is important to find out why.

We considered the 15 large SEP events that were not associated with the major eruptions as a separate group (Table 5), but we compiled flare and CME information for comparison with other tables. Note that the union of Tables 2-4 accounts for all the 59 eruptions. Similarly, the union of Tables 2 and 5 accounts for all the large SEP events in the study period.

For each event in Tables 2-5, we compiled the following information: Date and time of the associated flare and the flare size, flare location (Flare Loc.), flux rope locations (FR Loc.), solar B0 angle (inclination of the solar equator to the ecliptic), effective flux rope location after correcting for the B0 angle (Final Loc.), sky-plane speed of the associated CME ($V_{sky}$), and the peak speed ($V_{pk}$) of the fitted flux rope. Following the graduated cylindrical shell (GCS) model (Thernisien 2011) we fitted a flux-rope to the CMEs in the SOHO and STEREO images. According to the GCS model, the flux rope expands self-similarly with a circular front and conical legs of circular cross section. From the height-time measurements of the fitted flux rope, we obtained $V_{pk}$. The flux rope fit gives the heliographic coordinates of the CME, which may not coincide with the flare location if the CME moves non-radially. A large deviation between the flare and flux rope locations is indicative of the extent of non-radial CME motion. The deviation in the latitudinal direction is more important because of the stringent requirement of ecliptic distance discussed in Paper 1. The solar B0 angle also affects the latitudinal distance of the source region from the ecliptic. There are also additional information in the tables that are specific to each table.

**2.1 Major Eruptions Associated with Large SEP Events**
We first considered the 16 major eruptions associated with large SEP events listed in Table 2. These are potential sources of GLEs because they were associated with large SEP events and major flares/CMEs. Seven of these 16 events were analyzed Paper 1 and we included them here for completeness. In the last column of Table 2, we give the energy range (Max E) of the highest GOES energy channel in which a discernible SEP signal was detected. These observations were from the Energetic Proton, Electron and Alpha Detectors (EPEAD) on board the GOES 13 satellite. When there was a signal in the >700 MeV channel, the event was considered as a GLE event because it is likely to be observed by neutron monitors. For the 2012 May 17 GLE, SEPs were detected in the >700 MeV channel (see Paper 1).

The events in Table 2 were very energetic eruptions with $V_{pk}$ averaging to 2329 km/s, which is slightly greater than the average speed of GLE CMEs in cycle 23 (~2100 km/s). As column 10 in Table 2 shows, SEPs were detected at energies >700 MeV in only one case, the 2012 May 17 GLE. The CME speed was 1997 km/s, which was only ~14% below the average speed of Table-2 CMEs and the CME was well connected to Earth. Only two events (##11 and 13) had peak speeds well below the average: 1415 and 1626 km/s. In addition, the source locations (S23W06 and N14E11) indicate poorer connectivity that compounded the low speed. The 2013 May 22 (#14) event involved interacting CMEs. The peak speed of the SEP-producing CME was estimated to be 1881 km/s, which is ~19% below the average speed. The flux-rope fit was not accurate for this event because of the preceding CME. Since the source location was at W70, the sky-plane speed needs to be close to the true speed. Deprojecting the sky-plane speed (1466 km/s), we estimated the peak speed as 1560 km/s, suggesting that this was a slower event. The CME interaction signature was prominent in the dynamic spectra obtained by both Wind and STEREO radio receivers. It must be noted that all the >1000 pfu events cycle 24 involved interacting CMEs, including the 2013 May 22 event, which had a >10 MeV flux of 1660 pfu, second only to the 2012 March 7 event.

For events 3, 6 and 14, the source longitudes (E83, E31, and E64) were not in the traditional GLE longitude zone (>E15), although the CME speeds were very high. Therefore, the lack of GLEs is likely to be due to the poor longitudinal connectivity. The last event (#16) was the fastest (3121 km/s) in Table 2, yet it was not a GLE event most likely due to poor latitudinal connectivity (S15W29). In fact, the poor connectivity (C) seems to be the predominant factor in Table 2 because 9 events had poor latitudinal connectivity and two had poor longitudinal connectivity, accounting for 69% of the events. For event #13, the low speed was compounded by the marginal connectivity (N14E11).

The above analysis shows that there were only 3 well-connected high-speed events that are potential GLE candidates. One of them -- the 2012 May 17 eruption – was indeed a GLE event. The lack of GLEs in the remaining two events is certainly surprising. One of these events (2011 August 9) was already discussed in Paper 1: the CME was narrower than usual and the shock was weak, indicating an ambient medium of higher Alfven speed (Gopalswamy et al. 2008a,b). Events with an unfavorable ambient conditions are marked "A" in Table 2. The standoff distance of the shock structure is unusually large, which is the sign of a weak shock. The 2011 August 4 event (#1 in Table 2) was in a similar situation in that the corona in the nose part (around position angle 298º) was very dim, indicative of lower density and higher Alfven speed. Figure 1 shows the CME in the northwest quadrant as observed by SOHO/LASCO. While most of the CME was passing through a tenuous corona, the southern flank was interacting with a streamer, where the shock was likely to be stronger. However, the flank speed is expected to be smaller than the nose speed, so the shock strength may not be too high. Furthermore, the shock-streamer interaction region was not close to the ecliptic, so the latitudinal connectivity may not be favorable. In STEREO-Ahead (STA) view, this event was a limb event and we confirmed that the corona was tenuous in the nose part of the CME from COR2 images. The interplanetary type II burst observed by both Wind/WAVES and STA/WAVES was of narrow band, suggesting that only a small section of the shock was accelerating particles, consistent with the streamer interaction. In summary, we see that poor connectivity (11 C- events), low CME speed (3 V-events), and unfavorable ambient conditions (2 A-events) might have led to the lack of GLEs in the major eruptions in Table 2. Two events were double-counted under low speed and poor connectivity (C, V). This result is consistent with that in Paper 1 in that the poor latitudinal connectivity is the predominant factor that can explain the lack of GLEs in these major eruptions.

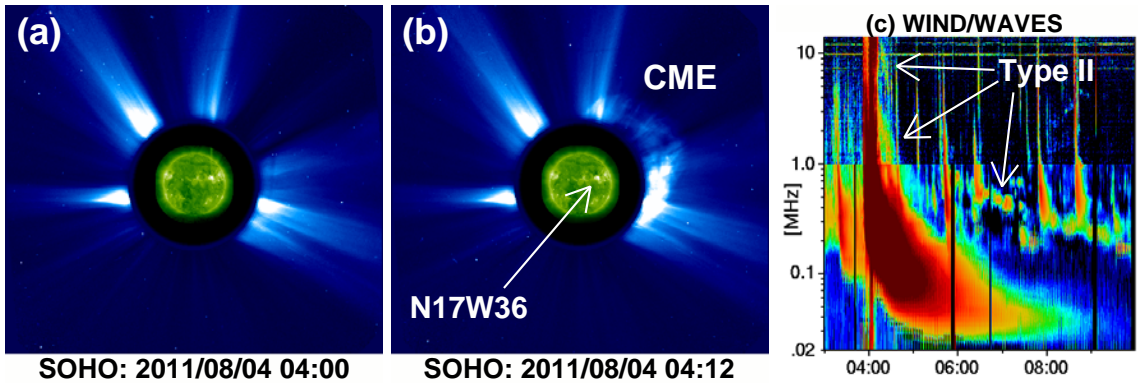

*Figure 1. SOHO/LASCO images of the pre-event corona (a) and the CME (b). In (a) and (b) an EUV image from SDO/AIA at 193 Å is superposed, which shows the solar source of the eruption (pointed by arrow). The Wind/WAVES radio dynamic spectrum shows the narrow band type II burst associated with the CME (c).*

In two cases in Table 2 (2012 January 27 and 2012 March 7) particles were detected in the 510-700 MeV channel, suggesting that these were "almost" GLE events. Both these events had poor latitudinal connectivity and, in addition, event 6 was poorly connected longitudinally (E27). In these two cases, it is highly likely that >700 MeV particles were accelerated, but they did not reach Earth in sufficient numbers to be detected as a GLE. The 2012 March 7 event had two CMEs in quick succession, which could not be seen distinct in the SEP time profile, so we combined them as one. The March 7 event was one of the three eastern SEP events (east of E15) in Table 2, the other two being 2011 September 22 (N09E89; 2474 km/s) and 2013 May 15 (N12E64; 2294 km/s). The CMEs were very fast (>2200 km/s) and the latitudes were in the right range. However, the >10 MeV event sizes were very small (35-41 pfu) and the events were observed only in the lowest energy channels. Fortunately, STB was well connected to these eastern events, so it was possible check the size of the SEP event from STB data.

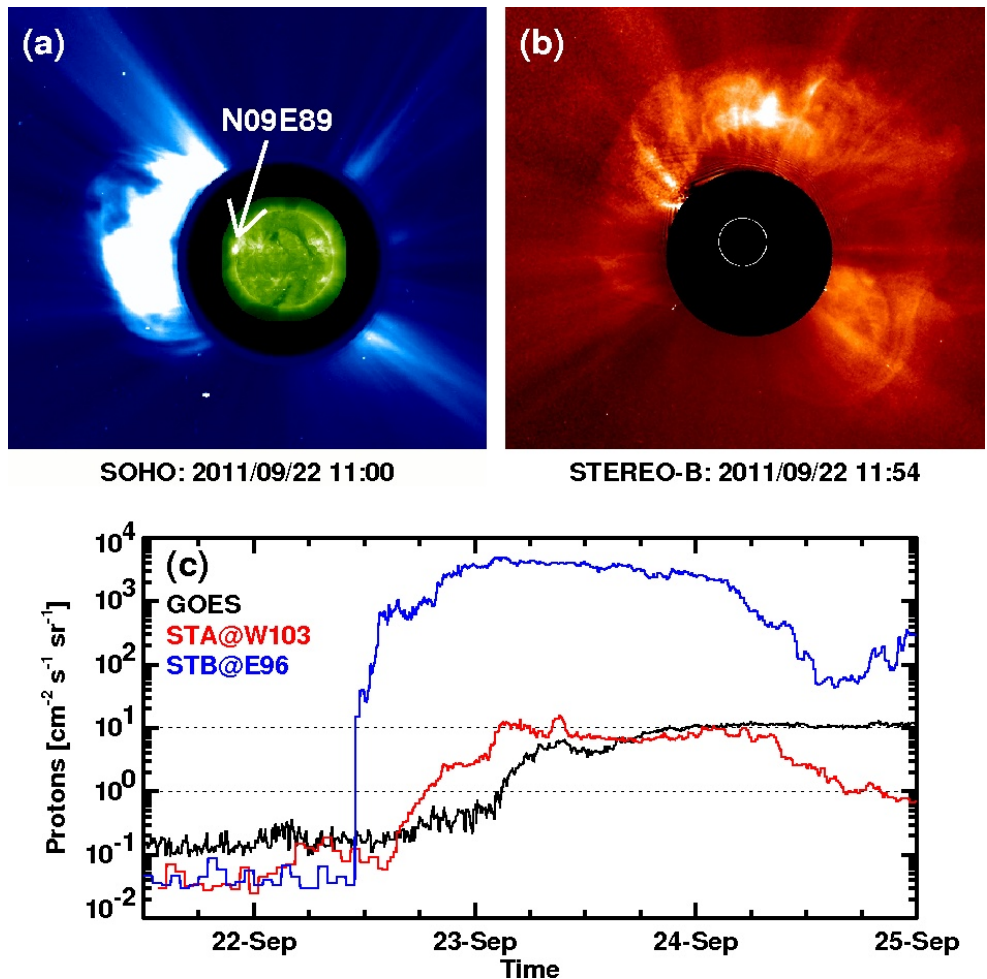

*Figure 2. The 2011 September 22 CME in SOHO/LASCO (a) and STB/COR2 (b) along with the >10 MeV flux from STB, STA, and GOES (c). An EUV image from SDO/AIA at 193 Å is superposed in (a) to show the solar source (pointed by arrow). The separation angle of STA and STB from Earth are given in (c).*

In order to get an estimate of the >10 MeV proton flux from STA and STB, we used the STEREO/IMPACT/HET data obtained in several energy channels from 13.6 to 100 MeV. We fitted a power law to the proton flux in these channels and derived the >10 MeV flux in the extrapolated range of 10 - 150 MeV. We neglected the flux of protons at energies > 150 MeV. Comparison between >10 MeV flux from GOES and STEREO/HET has shown close agreement when the STEREO spacecraft were very close to the Sun-Earth line. We applied this method to compare the >10 MeV flux from GOES, STA and STB for the 2011 September 22 event. LASCO observed the CME as an east-limb event (Fig. 2a). The CME was observed as a halo by STB/COR2 (Fig. 2b) because the source location was W07 in STB view (STB was located at E96 on this day). The CME was a back-sided halo (not shown) because the source location was at E192 in STA view (STA was at W103). Halo CMEs are regular CMEs that appear to surround the coronagraph occulting disk in sky-plane projection (see e.g., Howard et al 1982; Gopalswamy 2004, 2009; Gopalswamy et al. 2010a). The CME source was marginally well-connected to STB and the intensity of the >10 MeV particles was very high: ~5000 pfu. The event was also observed at Earth (GOES) and STA, but the intensity was rather low (~10 pfu). STA was located at W104, so the source was ~78º behind the west limb in STA view). It is remarkable that particles reached such widely separated locations around the Sun. Such multi-view observations are useful in deriving the longitudinal distribution of SEP intensity (Lario et al. 2013). The very high >10 MeV flux in the 2011 September 22 event points to the possibility of GeV particles accelerated in this event because there is a weak correlation between >10 MeV intensity and the GLE percentage (Gopalswamy et al. 2012). Unfortunately, STEREO HET does not observe protons above 100 MeV, so we do not know whether GeV particles were produced.

The fast CME (~2294 km/s) on 2013 May 15 CME originated from N11E64 and had a peak flux of ~40 pfu in the GOES >10 MeV channel. When we estimated the >10 MeV flux from the STB/HET data, we found that the event occurred during the decay phase of a previous event from the same region. It was barely observed as a distinct injection, so we could not determine the SEP event size. However, the background flux level was ~60 pfu, so it was certainly not a large event. More details on the previous events is discussed in the next subsection.

**2.2 Major Eruptions from Eastern Longitudes**
The large SEP events from the eastern hemisphere (Table 2) prompted us to examine all such eruptions originating from outside the GLE longitudes (east of E15). There were twenty four such eruptions (see Table 3). Three events (##3, 6, and 14) were also listed in Table 2 because of their association with large SEP events. We computed the >10 MeV flux from the STB/HET data as described in the previous subsection for the rest of the 21 eruptions to see if the flux exceeded 10 pfu. The results are shown in the last column of Table 3.

There were 10 events in Table 3 with high-speed CMEs (>2000 km/s), similar to most of the events in Table 2. STB was located in the range E82 (2010 November 6) to E114 (2013 November 8) for the eruptions in Table 2. Therefore, STB should have detected an SEP event because it was well connected to these 10 eruptions. Four of the 10 high-speed CMEs occurred when the background SEP intensity due previous events was high (HiB). Four of the remaining 6 eruptions produced very intense SEP events at STB with >10 MeV flux exceeding 1000 pfu (##03, 06, 20, 32). Even GOES did not observe this many high-intensity events over the whole of cycle 24. The 2011 September 22 event (#03) was the most intense in Table 3 with a flux of ~5000 pfu. The HiB events with fast CMEs all occurred following these high-intensity SEP events (see Table 3). The 2013 May 13 eruption at 01:53 UT (#31) resulted in an SEP event with only ~20 pfu. The second eruption (#32) on the same day had a 1000-pfu SEP event. The third and fourth eruptions (##33 and 14) occurred when the SEP background was high. Note that #32 followed #31 within 14 hours from the same source region and produced a much larger event (~1000 pfu). This is consistent with the scenario that CMEs preceded by wide CMEs from the same source region within 24 hours tend to produce high-intensity SEP events (Gopalswamy et al. 2004). The only high-speed event with no SEP association was the 2012 November 13 event (#29): neither GOES or nor STB detected an SEP event, but the CME was narrow and was not associated with a type II burst in the metric or longer wavelengths. We also confirmed that the CME was narrow in STA and STB coronagraphs, suggesting that it was a jet-like CME. Such narrow CMEs are known to be not associated with large SEP events (see e.g., Kahler et al. 2001).

There were three other large SEP events (## 34, 35, 36 in Table 3) at STB with CME speeds in the range 837 – 1531 km/s and one minor SEP event (#30). Five lower-speed CMEs occurred during high-SEP background, so we do not know if they caused a new injection of particles. The remaining five CMEs had no SEP association (denoted by "None" in the last column of Table 3, excluding the narrow CME in #29), but the CME speeds were in the range 419 to 1587 km/s, with an average of 864 km/s. The events in Table 3 demonstrate that the longitudinal connectivity and CME speed are important for SEP events and hence for the highest-energy particle events.

**2.3 Well-connected Major Eruptions without SEP Events**
The remaining 22 of the 59 eruptions had their source longitudes west of E15 (GLE longitudes), but were not associated with large SEP events, so they are also unlikely to produce GLE events. These events are listed in Table 4. We also checked if these eruptions were associated with minor SEP enhancements using the compound plots (involving GOES protons, CME height-time history, and GOES soft X-ray flux) available at the CDAW Data Center (e.g. for event #41 in Table 4: http://cdaw.gsfc.nasa.gov/CME_list/daily_plots/sephtx/2011_02/sephtx_20110215.png). From these plots, we estimated the peak proton flux and listed in the last column of Table 4. In 8 cases, there was no SEP enhancement at all (marked "None" in the last column). The average speed of these CMEs (930 km/s) was well below the average speed of SEP-associated CMEs (~1500 km/s). In 7 cases, there were minor events (<10 pfu) and the corresponding >10 MeV flux from GOES are shown in the last column of Table 4. These CMEs were slightly faster, ranging from 873 km/s to 1773 km/s with an average of 1343 km/s. The background was elevated but below 4 pfu in four events and the CME speeds were also low (average speed 1131 km/s). Thus we do not expect GLEs in these three categories. In three events (## 42, 48, and 49), the background was >10 pfu, so the existence of a large SEP event cannot be ruled out. The CME in event #42 was very slow, so it is unlikely to have caused an SEP event. The fastest CME (2157 km/s) in Table 4 was in event #49 (2012 March 10), but this CME occurred when the SEP background was very high (>100 pfu in the >10 MeV GOES channel). The latitudinal connectivity in this event was not good because of the large B0 angle (-7°.23): the flux rope location was N18W20, so the final location was N25W20. The CME in event #48 (2012 March 09) was also fast (1737 km/s), but the CME also occurred during a higher SEP background (>500 pfu in the >10 MeV GOES channel). The final flux rope location was from N12E03, so the latitudinal connectivity was in the right range, but the CME speed and longitudinal connectivity to Earth were not favorable for a GLE. The high background was due to the high-intensity event on March 7 (#6 in Table 2 and 3). Although we cannot rule out the existence of a large SEP event in these two eruptions, we did not find any GLE signature in the neutron monitor data.

When the two CMEs with high SEP background were excluded, the average speed of CMEs in Table 4 is ~1093 km/s, which is nearly half of the average speed of GLE CMEs. Therefore, it is not surprising that these eruptions were not associated with GLEs or even large SEP events. Ten of the events in Table 4 were outside the traditional GLE longitudes (W20-W90), eight of which were under the "None" category. The two well-connected events (##58 and 59) without SEP signatures had very low CME speeds (749 and 548 km/s). If we consider the best-connected events in Table 4 (## 46-47, 53-55, 59), we see that the speeds were in the range 548 to 1773 km/s, with an average of 1120 km/s. The second fastest (1773 km/s) event in Table 3 (#54 - 2013 October 28) was preceded by a fast CME (953 km/s) from an eruption (#53) that occurred only ~3 h before #54 from the same active region. Wind/WAVES dynamic spectrum showed an interaction signature during 6:00 to 8:00 UT and the SEP flux started increasing during this time. This seems to be a case of the ambient medium effect, but in enhancing the chance of SEP acceleration. Excluding the cases of high particle background and CME interaction, we conclude that the CME speed is the primary factor contributing to the lack of SEP events (and hence GLE events) in these eruptions.

**3. Large SEP Events Associated with Weaker Eruptions**
In cycle 24, there were 31 large SEP events as of January 31, 2014. Only 16 of these overlapped with the list of major eruptions (≥M5.0) considered above. The remaining 15 large SEP events were associated with smaller eruptions in terms of the flare size (see Table 5). GLE events are always associated with large SEP events and fast CMEs, even though occasionally they may be associated with C-class flares. Therefore, we considered the characteristics of the remaining 15 large SEP events as listed in Table 5 to see why there were no GLE events. The columns in Table 5 are the same as in Table 2.

Five of these large SEP events were due to eruptions from behind the west limb, so their soft X-ray flare sizes were unknown. Therefore, only 10 SEP events in Table 5 were truly from weaker eruptions with flare sizes ranging from C1.2 to M3.7. The two events with the weakest soft X-ray enhancements (##5, 14 in Table 5) were associated with eruptive prominences (Gopalswamy et al. 2014, under preparation). These eruptions had clear post-eruption arcades that corresponded to the C-class GOES flares. The CMEs in the 10 eruptions were generally fast with an average speed of 1724 km/s. Recall that the average speed of SEP-associated CMEs in Table 2 was ~2300 km/s, so the average speed of CMEs in the weaker eruptions was only ~26% smaller. Table 5 also shows that the maximum intensity of SEP events in frontside events is only 200 pfu, compared to 3000 pfu in Table 2. There were also five >100 pfu events in Table 2 compared to just 2 in Table 5. The intensity difference is consistent with the well-known correlation between SEP intensity and CME speed (see e.g., Kahler 2001). The high speed of frontside CMEs in Table 5 also suggest that the CME speed is more important than the flare size.

Only one of the 10 front-side eruptions in Table 5 had a peak speed exceeding 2000 km/s: the 2011 March 7 CME was very fast (2660 km/s) and associated with an M3.7 flare. The flare location was N24W59, but the flux rope location was N32W58. The B0 angle was also not favorable (-7º.3), so the final location of the CME became N39W58. This CME was longitudinally well connected to Earth, but the ecliptic distance was 3 times larger than the average value of cycle-23 GLE CMEs. The type II radio burst was observed over a wide range of wavelengths, indicating a strong shock. We suggest that high energy particles might have been accelerated at the shock nose, but they were not observed at Earth because the shock nose was not connected to Earth.

All the five backside CMEs in Table 5 had peak speeds around 2000 km/s or higher and hence relevant for the production of GLEs (## 3, 6, 9, 14, and 15 in Table 5). These events are frontside events for STA, so we computed the >10 MeV flux from the STA/HET data as described in section 2.2. The two fastest events on 2012 May 26 (#6 - 2623 km/s) and 2012 July 23 (#9 – 2621 km/s) resulted in very intense SEP events in STA. For these two events, the latitudinal connectivity was good (N12 for #6 and N00 for #9), but the longitudinal connectivity is poor (~25º and ~45º behind the west limb, respectively). Even though the speeds were the same, the GOES proton data showed that particles were detected only up to the 15-40 MeV channel for #6, while up to the 165-500 MeV energy channel for #9 (see Table 5). The CMEs were heading toward STA in both cases (source longitude was W00 for #6 and W14 for #9 in STA view). Therefore, both CMEs resulted in huge ESP events.

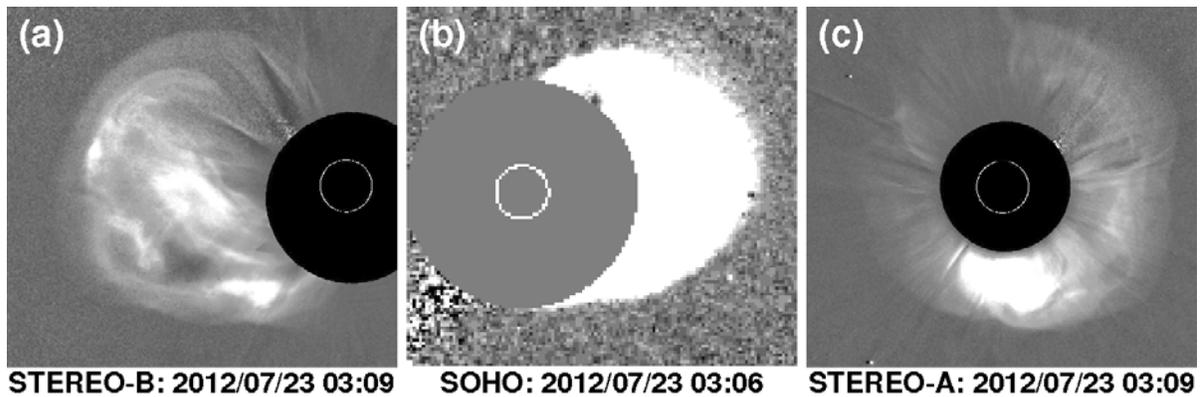

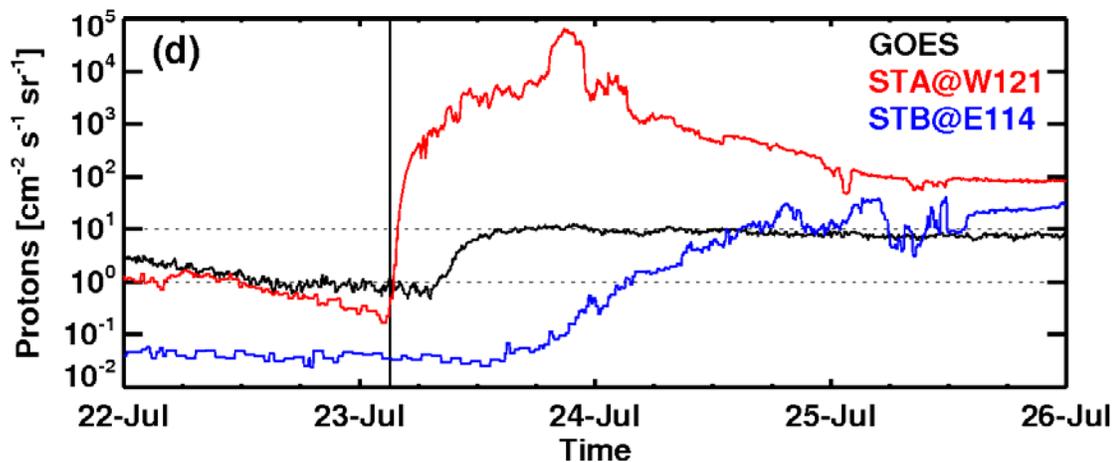

Figure 3. Three snapshots of the 2012 July 23 backside event taken around the same time: (a) STB, (b) SOHO/LASCO, and (c) STA. The CME was heading toward STA, so it appears as a symmetric halo. (d) >10 MeV flux at STA, STB, and GOES. The separation angle of STA and STB from Earth are given in (d).

Figure 3 shows the 2012 July 23 CME from three views and the >10 MeV particle flux. The CME appeared as a symmetric halo in STA view, because it was heading toward that spacecraft. The CME appeared as asymmetric halos in STB and LASCO FOV at later times. The >10 MeV SEP flux at STA was ~5000 pfu, with an order of magnitude larger ESP event (~$6.5 \times 10^4$ pfu). These fluxes may be slightly overestimated because of a spectral turnover at the lowest energies, but still they are larger than the largest of cycle 23 (Gopalswamy et al 2005; Mewaldt et al. 2005). The STA spacecraft (located at W121) was longitudinally better connected to the source region (W135) than Earth and hence observed the extreme SEP event (Russell et al. 2013; Baker et al. 2013). The event onset was delayed by ~4 h in GOES and by ~8 h in STB due to connectivity. The STB event was only slightly larger than the GOES event. This event also demonstrates that far-side events can contribute significantly to the SEP flux at Earth (Mewaldt et al. 2013). As noted before, STEREO/HET does not detect protons at energies >100 MeV, so we cannot say whether this was a GLE-type event on the backside. The flux values for slightly lower for the 2012 May 26 event with a >10 MeV SEP flux of ~1500 pfu and ESP flux of ~$10^4$ pfu. Even #3 (2011 March 21) was fast but the latitudinal distance from the ecliptic was quite large (N33). The eruption was 35° behind the west limb. Since STA was at W88 on this day, the eruption was well connected to STA (W47 in STA view). The >10 MeV flux was accordingly large (~1500 pfu). Eruptions not connected to Earth but associated with >1000 pfu SEP events in Tables 3 and 5 were detected primarily because of the availability of STEREO observations.

The last two events in Table 5 (##14 and 15) had high speeds and at the right latitudes. Event #14 (2013 December 28) was from N02W127, so it was an eastern event (E23) in STA view. STB was located at E151, so the eruption was from W278 (or E82) in STB view, so the two STEREO spacecraft detected only a minor event (~1 pfu). This event is somewhat similar to the 2011 August events in Table 2. The type II burst was extremely weak, suggesting that the shock was very weak. This seems to be a case of high Alfven speed in the ambient medium. Finally, event #15 was indeed a GLE event as evidenced by particles found in the highest energy channel (>700 MeV). South Pole Neutron Monitors also detected this event as a GLE (Thakur et al. 2014) but of low intensity (~2.5% above the background). All the factors were favorable for this eruption to be a GLE event: latitudinally well connected, only slightly behind the limb, and high CME speed (~2200 km/s, but some estimates put it as low as 1660 km/s – see Thakur et al. 2014). The GLE event is longitudinally better connected (only ~12° behind the limb) among all the backside events in Table 5. This is one of the weakest SEP events (>10 MeV flux was ~40 pfu) to be associated with a GLE event. The >10 MeV SEP flux at STA (located at W150) and STB (located at E152) were 1 and 0.3 pfu, respectively.

Figure 4 shows the 2014 January 6 CME along with the proton intensity in several channels from 9 MeV to >700 MeV. Also shown for comparison is the 2014 January 7 CME and the proton intensity (#16 in Table 2). Although the January 7 CME was of higher speed and slightly better connected longitudinally, its latitudinal connectivity to Earth was poorer. This is clear from the LASCO image of the CME in Fig. 4, which shows that the CME nose was well below the ecliptic plane and hence was not connected to Earth. The flare latitude was nearly the same in both cases (S15W108 and S15W11 for the January 6 and January 7 flares, respectively), but the effective location after taking into account of non-radial motions and B0 angle left the January 6 event latitudinally better connected than the January 7 CME. The reason for the non-radial motion of the 2014 January 7 CME seems to be due to a combination of deflection due to a large coronal hole in the northeast quadrant (similar to cases in Gopalswamy et al. 2009) and the large arcade in the active region that did not participate in the eruption (see e.g., Sterling et al. 2011). This comparison also supports the suggestion that the highest energy particles may be accelerated in the nose part of the CME, but to detect them, the nose has to be well connected to the observer.

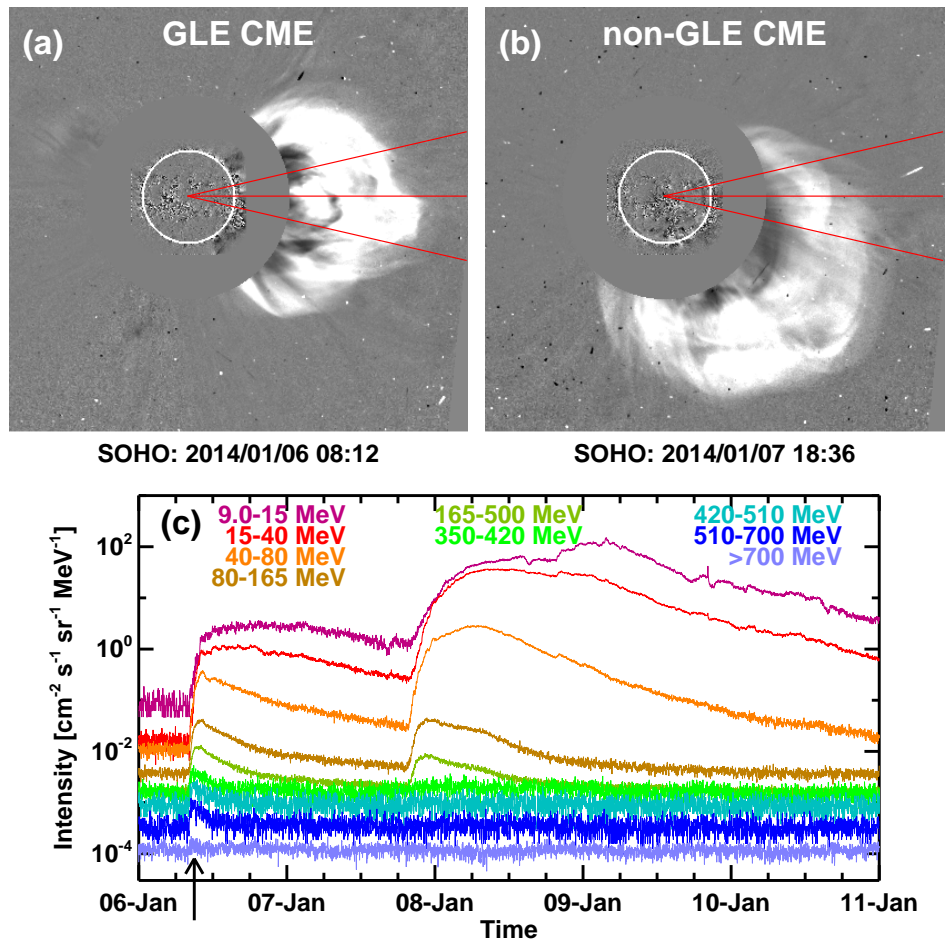

Figure 4. Snapshots of the GLE CME on 2014 January 6 (a) and an equally energetic CME on 2014 January 7 that did not produce a GLE (b). The lines superposed on the coronagraph images delineate the region (±13º from the ecliptic) into which the CME nose needs to be for a GLE event to be detected at Earth. Proton intensity plots for the two events (c) show that the CME (a) had energetic particles in the >700 MeV channel (pointed by arrow), while the CME (b) had particles only up to the 165-500 MeV channel.

Events 4, 5, and 11 were the three best-connected events in Table 5 and associated with large SEP events. However, the CME speeds (1680, 1187, and 1479 km/s, respectively) were below the average speed of the CMEs in Table 5 (1724 km/s). In these three cases, we can confidently say that the highest energy particles were not accelerated in sufficient numbers to be detected as GLEs. This is probably true for all the other front-sided events in Table 5 because their average speed is well below the speed of GLE CMEs. The backside events all had high speeds appropriate for GLE events, but the poor connectivity to Earth did not permit them to be observed as GLEs. Unfortunately, the STEREO spacecraft do not have particle detectors that observe the highest energy particles. Thus we conclude that most of the SEP events in Table 5 did not become GLE events because of the same three factors discussed before: CME speed, latitudinal/longitudinal connectivity, and ambient conditions. This is in contrast to the events in Table 2, where the connectivity was the predominant factor. The main factor contributing to the lack of GLEs seems to be the CME speed.

**3. Discussion**

One of the interesting results of this study is that the flare size is not a good indicator of an SEP event, but the CME speed is. In Table 1 we saw that only 16 of the 59 major eruptions (or 27%) were associated with large SEP events. Since eastern events are less likely to produce SEP events at Earth, we considered 35 western (>E15) eruptions and found only 13 out of 35 or 37% were associated with large SEP events. This fraction was not significantly different from the fraction of large SEP events associated with weaker eruptions (flare size <M5.0): 10 out of the 26 frontside SEP events (or 38%) in cycle 24. On other hand, the CME speed was consistently high for both the major eruptions in Table 2 (2329 km/s) and the weaker eruptions in Table 5 (1724 km/s). It is true that the average speed of CMEs was slightly smaller for the set of weaker eruptions, but well above the speed required for driving a shock near the Sun that can accelerate energetic particles. Thus the observations lend support to the idea that CME shocks are the primary source of large SEP events.

Gopalswamy (2012) reported that the SEP-associated CMEs in solar cycle 24 had a higher average speed and larger halo fraction than those in cycle 23 over the corresponding epoch. A recent investigation confirmed that this is due to the altered state of the heliosphere with reduced total pressure and magnetic field (Gopalswamy et al. 2014; see also McComas et al. 2013). In particular, the decrease in the ambient magnetic field reduces the acceleration efficiency, which might explain the paucity of GLEs in cycle 24. This is consistent with the fact that there were 7 GLEs in the first 62 months of cycle 23 (similar to our study interval), while there were only 2 in cycle 24. If we extend the period by another 5 months (to the end of June 2014), cycle 24 still had only 2 GLEs vs. 9 in cycle 23. Moreover, the average speed of the 7 cycle-23 GLE CMEs was only 1620 km/s, which is smaller by 31% compared to the average speed of CMEs in Table 2. The two GLE CMEs of cycle 24 were also faster than 1620 km/s. Even the average speed of SEP CMEs over the first five years of cycle 23 was smaller than that in cycle 24 (Gopalswamy 2012). The higher CME speed required in cycle-24 CMEs to accelerate particles to the same or lower energies is also indicative of a lower particle acceleration efficiency in cycle 24. The overall reduction in the particle-acceleration efficiency will naturally decrease the frequency of occurrence of GLE events. For two events in Table 2 (2011 August 4 and 11) and one event in Table 5 (2011 March 07), the speed and connectivity were appropriate for GLEs, yet they lacked GLEs. In these cases, the overall reduction in acceleration efficiency seems to be compounded by the local variation in the ambient medium. It was noted before that Alfven speed can vary by a factor up to 4 from event to event (Gopalswamy et al. 2008a,b; 2010b).

An additional factor that became clear from Table 2 is that many high-speed CMEs had effective source locations at large distances from the ecliptic. These represent poor latitudinal connectivity. While discussing magnetic connectivity to Earth, it is customary to consider longitudinal connectivity. The latitudinal connectivity for GLEs was introduced only recently (Gopalswamy et al. 2013; Gopalswamy and Mäkelä 2014), which seems to be different for non-GLE SEP events (Dalla and Agueda 2010 and references therein). The latitudinal connectivity is affected mainly by non-radial motion of CMEs and the solar B0 angle. The non-radial motion of CMEs is either due to inherent imbalance in the active region (Sterling et al. 2011) or due to deflection by coronal holes (Gopalswamy et al. 2009c; Xie et al. 2013; Mäkelä et al. 2013). The solar B0 angle can also improve or decrease the latitudinal connectivity by a maximum of ~7°.23, whereas the effect of non-radial motion can be much larger. In fact, the connectivity for the 2012 May17 GLE improved when a coronal hole deflected the CME toward the ecliptic (see Fig. 1 in Gopalswamy and Mäkelä 2014). Paper 1 reported that the average latitudinal distance of CME source regions in GLE events is 13° from the ecliptic. This suggests that the highest-energy particles may be accelerated near the nose of the shock and the nose needs to be connected to Earth to be detected a GLE. On the other hand, lower energy particles may be accelerated over a larger extent of the shock as suggested by the non-GLE SEP events in Table 2 that have poor connectivity in latitude or longitude.

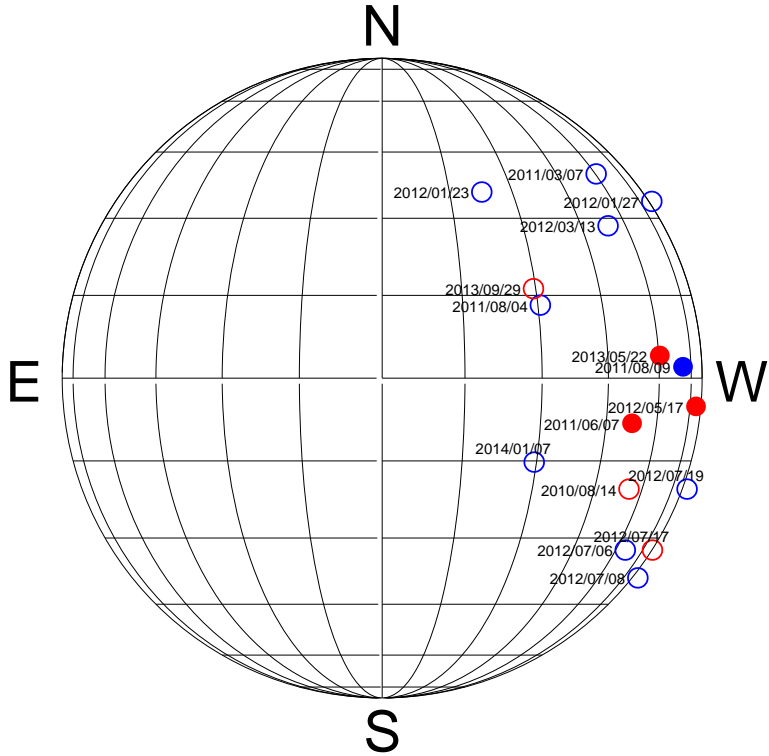

*Figure 5. Heliographic locations of the CME flux ropes of the 16 fast (>1500 km/s) eruptions from traditional GLE longitudes (W20 – W90). The flux rope locations obtained from the Graduated Cylindrical Shell model has been corrected for solar B0 angles (see text). The CMEs are further distinguished based on their speeds (≥2000 km/s in blue and 1500 km/s < speed <2000 km/s in red) and latitudinal distance from the ecliptic (filled circle within 13º and open circles at or above 13º).*

Figure 5 summarizes the major points of the paper by focusing on fast (>1500 km/s) CMEs originating from the traditional GLE longitudes (W20-90) and associated with large SEP events. There were 16 such fast events extracted from Tables 2 and 5 that had high enough speeds and longitudinally well connected to Earth as the best candidates for accelerating GeV particles. The CMEs are also color coded to distinguish the ≥2000 km/s CMEs (blue) from the <2000 km/s CMEs (red). In addition, CMEs with effective source locations within 13º from the ecliptic (filled circles) and outside of this latitudinal distance (open circles) are distinguished. We see that all but one of the ten ≥2000 km/s CMEs were poorly connected latitudinally. On the other hand, half of the six <2000 km/s CMEs had poor latitudinal connectivity. There were only four best-connected CMEs that in the right latitudinal range, of which only one was a GLE. While we cannot say whether or not the highest energy particles were accelerated in the 12 CMEs with poor latitudinal connectivity, we can definitely say that there were no GeV particles in the four best-connected fast CMEs. The only way the other three events did not produce a GLE event is that the shocks were weak due to the state of the heliosphere discussed above.

A final remark on the shock geometry in order. By independent techniques, Reames (2009) and Gopalswamy et al. (2012; 2013b) have shown that the shock first forms at a heliocentric distance of ~1.5 Rs (based on type II radio burst onset) and it takes another 1.5 Rs before the highest energy particles are released from the shock. When the leading edge of the CME is at a distance of 3 Rs, the shock nose is likely to be just above the source surface and the local shock geometry is quasi parallel. Since the critical Mach number of quasi-parallel shocks (~1.5) is much lower than that of quasi-perpendicular shocks (~2.9), the former are likely to be supercritical at a heliocentric distance of about 3 Rs, where the Alfven speed peaks (Mann et al. 1999; Gopalswamy et al. 2001; Mann et al. 2003). For a 2000 km/s CME, the Alfvénic Mach number is expected to be around 3, which is definitely supercritical only for a quasi-parallel shock.

## 4. Summary and Conclusions

We studied all the major solar eruptions that occurred on the frontside of the Sun during solar cycle 24 until 2014 January 31. Fifty nine eruptions were selected based on the criteria that the soft X-ray flare size must be ≥M5.0 and accompanied by a CME. We examined the association of these eruptions with large SEP events and GLE events. We found that only 16 of the 59 eruptions (or 27%) were associated with large SEP events, including the 2012 May 17 GLE. There were 31 large SEP events during the study period, which means only about half of the large SEP events were associated with major eruptions. Of the remaining 15 large SEP events, five originated from behind the limb, so we do not know their true flare size. One of these five backside events was a GLE (2014 January 6). The remaining 10 SEP events were associated with weaker eruptions (soft X-ray flare size <M5.0), although the CME speeds were high enough to drive shocks that accelerated particles. Major eruptions that did not have a large SEP event were either poorly connected longitudinally (east of E15) or the CME speed was well below the typical speed of GLE CMEs. We also examined the reason for the paucity of GLE events during solar cycle 24. Most of the eruptions with high CME speed (>2000 km/s) had poor connection to Earth (latitude and /or longitude), so the highest energy particles may not have reached Earth, even though they may have been accelerated. There were a few eruptions with high CME speed and good connection, yet lacked GLEs. In these cases, the ambient medium is likely to have played a role either due to local high Alfven speed coupled with the reduced acceleration efficiency in the heliosphere with a weaker magnetic field. Many well-connected SEP events also did not have GLEs mainly because of the lower CME speed. Thus we conclude that three major factors contribute to the lack of GLEs in SEP events during cycle 24: (i) poor connectivity in latitude and/or longitude, (ii) lower CME speed than the typical GLE CME speeds, and (iii) ambient medium with unfavorable conditions for particle acceleration (locally or due to the overall reduction in acceleration efficiency due to the weak solar cycle). These are in addition to other factors such as preconditioning by preceding eruptions (Gopalswamy et al. 2003).

**Authors' contributions**
NG planned and led the study, interpreted the results, and drafted the manuscript. SA compiled the list of eruptive events. HX performed flux rope fitting of the events considered for the study. SY prepared the figures. PM checked the data analysis and interpretation. All authors critically read the manuscript and approved the final version.

**Acknowledgments:** This work was supported by NASA's LWS and TR&T programs. This work benefitted greatly from the open data policy of NASA.

Table 1. An overview of the major eruptions of cycle 24

| | |
|---|---|
| Major flares (≥M5.0) | 69 |
| Confined Flares (No CMEs) | 9 |
| Flares with no LASCO CME data | 1 |
| # Events for this study | 59 |
| # Eastern Eruptions (≤E15) | 24 |
| # Large SEP Events in SC 24 | 31 |
| # Eruptions with SEP Events | 16/59 (27%) |
| # Western Eruptions with SEPs | 13/35 (37%) |
| # SEP Events with <M5 Flares | 10/31 (32%) |
| #SEP Events in Backside Eruptions | 5/31 (16%) |

Table 2. Major eruptions associated with large SEP events in cycle 24

| No. | Flare Date & Time | Flare Size | Flare Location | FR Location | B0 deg | Final Location | $V_{sky}$ km/s | $V_{pk}$ km/s | Max E MeV | SEP[c] GOES | Reason[d] |
|---|---|---|---|---|---|---|---|---|---|---|---|
| 1 | 2011/08/04 03:41 | M9.3 | N19W36 | N19W30 | +6.0 | N13W30 | 1315 | 2450 | 165-500 | 96 | A,C |
| 2[a] | 2011/08/09 07:48 | X6.9 | N17W69 | N08W68 | +6.3 | N02W68 | 1610 | 2496 | 350-420 | 26 | A |
| 3 | 2011/09/22 10:29 | X1.4 | N09E89 | N05E83 | +7.1 | S02E83 | 1915 | 2474 | 80-165 | 35 | C |
| 4 | 2012/01/23 03:38 | M8.7 | N28W21 | N30W22 | -5.3 | N35W22 | 2175 | 2150 | 165-500 | 3000 | C |
| 5[a] | 2012/01/27 17:37 | X1.7 | N27W51 | N27W82[b] | -5.6 | N33W82 | 2508 | 2625 | 510-700 | 800 | C |
| 6 | 2012/03/07 00:02 | X5.4 | N17E27 | N18E31 | -7.3 | N25E31 | 2684 | 2987 | 510-700 | 1500 | C |
| 7[a] | 2012/03/13 17:12 | M7.9 | N17W66 | N21W52 | -7.2 | N28W52 | 1884 | 2333 | 420-510 | 500 | C |
| 8[a] | 2012/05/17 01:25 | M5.1 | N11W76 | S07W76 | -2.4 | S05W76 | 1582 | 1997 | >700 | 255 | ---- |
| 9[a] | 2012/07/06 23:01 | X1.1 | S13W59 | S29W62 | +3.4 | S32W62 | 1828 | 2464 | 165-500 | 25 | C |
| 10[a] | 2012/07/08 16:23 | M6.9 | S17W74 | S34W88 | +3.7 | S38W88 | 1495 | 2905 | 165-500 | 18 | C |
| 11 | 2012/07/12 15:37 | X1.4 | S15W01 | S19W06 | +4.1 | S23W06 | 885 | 1415 | 165-500 | 96 | C,V |
| 12[a] | 2012/07/19 04:17 | M7.7 | S13W88 | S15W88 | +4.7 | S20W88 | 1631 | 2048 | 165-500 | 70 | C |
| 13 | 2013/04/11 10:55 | M6.5 | No9E12 | N08E11 | -5.9 | N14E11 | 861 | 1626 | 165-500 | 114 | C,V |
| 14 | 2013/05/15 01:25 | X1.2 | N12E64 | N12E64 | -2.6 | N15E64 | 1366 | 2294 | 40-80 | 41 | C |
| 15 | 2013/05/22 13:08 | M5.0 | N15W70 | N02W59 | -1.8 | N04W59 | 1466 | 1881 | 350-420 | 1660 | V |

| 16 | 2014/01/07 18:04 | X1.2 | S15W11 | S19W29 | -3.7 | S15W29 | 1774 | 3121 | 350-420 | 1000 | C |

<sup>a</sup>From Paper 1. <sup>b</sup>The flux rope location has been revised slightly from Paper 1. <sup>c</sup>Peak >10 MeV flux in pfu; <sup>d</sup>The factors that contribute to the lack of GLEs: A – Ambient medium, C – connectivity in latitude and/or longitude, V – speed.

Table 3. Major Eruptions of cycle 24 poorly connected to Earth

| No | Date & Time | Size | Flare Loc | FR Loc | B0 | Final Loc | $V_{sky}$ | $V_{pk}$ | SEP[b] |
|---|---|---|---|---|---|---|---|---|---|
| 17 | 2010/11/06 15:27 | M5.4 | S19E58 | S19E57 | +3.8 | S23E57 | 178 | 583 | None |
| 18 | 2011/07/30 02:04 | M9.3 | N14E35 | N04E30 | +5.7 | S02E30 | 361 | 419 | None |
| 03 | 2011/09/22 10:29[a] | X1.4 | N09E89 | N05E83 | +7.1 | S02E83 | 1915 | 2474 | 5000 |
| 19 | 2011/09/24 09:21 | X1.9 | N12E60 | N00E59 | +7.0 | S07E59 | 1936 | 2975 | HiB |
| 20 | 2011/09/24 12:33 | M7.1 | N10E56 | N06E52 | +7.0 | S01E52 | 1915 | 2228 | 1000 |
| 21 | 2011/09/24 20:29 | M5.8 | N13E52 | N25E48 | +7.0 | N18E48 | 356 | 470 | HiB |
| 22 | 2011/09/25 04:31 | M7.4 | N11E47 | S12E31 | +7.0 | S19E31 | 788 | 1540 | HiB |
| 23 | 2012/03/05 03:17 | X1.1 | N17E52 | N34E51 | -7.3 | N41E51 | 1531 | 1628 | HiB |
| 06 | 2012/03/07 00:02 | X5.4 | N17E27 | N18E31 | -7.3 | N25E31 | 2684 | 2987 | 1500 |
| 24 | 2012/03/07 01:05[a] | X1.3 | N25E26 | S08E08 | -7.3 | S01E08 | 1825 | 2147 | HiB |
| 25 | 2012/07/28 20:44 | M6.1 | S25E54 | S17E50 | +5.5 | S23E50 | 420 | 1134 | HiB |
| 26 | 2012/08/18 00:24 | M5.5 | N19E86 | N39E88 | +6.8 | N32E88 | 986 | 1587 | None |
| 27 | 2012/10/20 18:05 | M9.0 | S13E79 | S07E79 | +5.5 | S13E79 | 330 | 799 | None |
| 28 | 2012/10/23 03:13 | X1.8 | S13E60 | S14E69 | +5.2 | S19E69 | 636 | 933 | None |
| 29 | 2012/11/13 01:58 | M6.0 | S25E46 | S34E49 | +3.1 | S37E49 | 851 | 2383 | None |
| 30 | 2013/05/03 17:24 | M5.7 | N16E81 | S24E89 | -4.0 | S20E89 | 858 | 1295 | 1 |
| 31 | 2013/05/13 01:53 | X1.7 | N11E90 | N11E101 | -2.9 | N14E101 | 1270 | 2318 | 20 |
| 32 | 2013/05/13 15:48 | X2.8 | N11E85 | N11E94 | -2.9 | N14E94 | 1850 | 2889 | 1000 |
| 33 | 2013/05/14 00:00 | X3.2 | N08E77 | N13E86 | -2.8 | N16E86 | 2625 | 2963 | HiB |
| 14 | 2013/05/15 01:25[a] | X1.2 | N12E64 | N13E74 | -2.7 | N16E74 | 1366 | 2294 | HiB |
| 34 | 2013/10/25 07:53 | X1.7 | S08E73 | S03E76 | +5.1 | S08E76 | 641 | 1531 | 20 |
| 35 | 2013/10/25 14:51 | X2.1 | S06E69 | N02E72 | +5.1 | S03E72 | 1221 | 1384 | 60 |
| 36 | 2013/11/05 22:07 | X3.3 | S12E46 | S30E46 | +4.0 | S34E46 | 611 | 837 | 60 |
| 37 | 2013/11/08 04:20 | X1.1 | S14E15 | S06E15 | +3.6 | S10E15 | 600 | 846 | HiB |

<sup>a</sup>These events were also in Table 2 because they were associated with a large SEP event at Earth; <sup>b</sup>>10MeV SEP flux observed at STB as computed from the HET data.

Table 4. List of M5 or larger flares non-SEP west of E15 (23 events)

| No | Flare Date & Time | Flare Size | Flare Location | FR Location | B0 Deg. | Final Location | $V_{sky}$ km/s | $V_{pk}$ km/s | SEP[a] GOES |
|---|---|---|---|---|---|---|---|---|---|
| 38 | 2010/02/07 02:20 | M6.4 | N21E10 | S17E11 | -6.4 | S17E11 | 421 | 847 | None |
| 39 | 2010/02/12 11:19 | M8.3 | N26E11 | N11E07 | -6.7 | N18E07 | 509 | 1056 | None |
| 40 | 2011/02/13 17:28 | M6.6 | S20E04 | S15E03 | -6.7 | S08E03 | 373 | 847 | None |
| 41 | 2011/02/15 01:44 | X2.2 | S20W10 | S16W11 | -6.8 | S09W11 | 669 | 1441 | 2 |
| 42 | 2011/03/08 10:35 | M5.3 | S17W86 | S25W87 | -7.3 | S18W87 | 398 | 491 | 30 HiB |
| 43 | 2011/08/03 13:17 | M6.0 | N16W30 | N17W14 | +5.9 | N11W14 | 610 | 1525 | 1 |
| 44 | 2011/09/06 01:35 | M5.3 | N14W07 | N20W17 | +7.2 | N13W17 | 782 | 873 | 2 |
| 45 | 2011/09/06 22:12 | X2.1 | N14W18 | N20W19 | +7.2 | N13W19 | 575 | 1474 | 8 |
| 46 | 2011/09/07 22:32 | X1.8 | N14W28 | N20W34 | +7.3 | N13W34 | 792 | 1240 | 3 HIB |
| 47 | 2011/09/08 15:32 | M6.7 | N14W40 | N20W46 | +7.3 | N13W46 | 214 | 782 | 1 HiB |
| 48 | 2012/03/09 03:22 | M6.3 | N15W03 | N05E03 | -7.2 | N12E03 | 950 | 1737 | 500 HiB |
| 49 | 2012/03/10 17:15 | M8.4 | N17W24 | N18W20 | -7.2 | N25W20 | 1296 | 2157 | 100 HiB |
| 50 | 2012/07/02 10:43 | M5.6 | S17E08 | S33E01 | +3.0 | S36E01 | 313 | 1314 | None |
| 51 | 2013/06/07 22:11 | M5.9 | S32W89 | S34W91 | +0.1 | S34W91 | 887 | 1189 | None |
| 52 | 2013/10/24 00:21 | M9.3 | S10E08 | S10E07 | +5.1 | S15E07 | 619 | 887 | None |
| 53 | 2013/10/28 01:41 | X1.0 | N04W66 | N07W76 | +4.7 | N02W76 | 806 | 953 | <1 |
| 54 | 2013/10/28 04:32 | M5.1 | N08W71 | N10W75 | +4.7 | N05W75 | 928 | 1773 | 4 |
| 55 | 2013/10/29 21:42 | X2.3 | N05W89 | N12W91 | +4.7 | N07W91 | 1169 | 1421 | 4 HiB |
| 56 | 2013/11/10 05:08 | X1.1 | S14W13 | S31W31 | +3.4 | S34W31 | 761 | 1081 | 1 HiB |
| 57 | 2013/11/19 10:14 | X1.0 | S14W70 | S29W71 | +2.4 | S31W71 | 776 | 1364 | 4 |
| 58 | 2013/12/31 21:45 | M6.4 | S15W36 | S30W36 | -2.9 | S27W36 | 408 | 749 | None |
| 59 | 2014/01/01 18:40 | M9.9 | S14W47 | S15W48 | -3.0 | S12W48 | 317 | 548 | None |

[a]Information on SEPs from GOES proton data. The numbers denote the proton flux (in pfu). "None" indicates no enhancement above the background. HiB – High background level indicated by the preceding number in pfu.

Table 5: Large SEP events of cycle 24 associated with weaker flares (<M5.0)

| No | Flare Date & Time | Flare Size | Flare Location | FR Location | B0 deg | Final Location | $V_{sky}$ km/s | $V_{pk}$ km/s | Max E MeV | SEP[b] GOES | SEP[c] STA (B) |
|---|---|---|---|---|---|---|---|---|---|---|---|
| 1 | 2010/08/14 09:38 | C4.4 | N12W56 | S13W54 | +6.6 | S20W54 | 1205 | 1658 | 40-80 | 14 | <0.1 |
| 2 | 2011/03/07 19:43 | M3.7 | N24W59 | N32W58 | -7.3 | N39W58 | 2125 | 2660 | 40-80 | 50 | 200 |
| 3 | 2011/03/21 02:00 | back | N24W129 | N26W125 | -7.0 | N33W125 | 1341 | 2022 | 80-165 | 14 | 1500 |
| 4 | 2011/06/07 06:16 | M2.5 | S21W54 | S08W51 | +0.1 | S08W51 | 1255 | 1680 | 350-420 | 72 | HiB |
| 5 | 2011/11/26 06:09 | C1.2 | N08W49 | N10W47 | +1.5 | N08W47 | 933 | 1187 | 40-80 | 80 | 10 |
| 6 | 2012/05/26 20:40 | back | N10W126 | N11W115 | -1.4 | N12W115 | 1966 | 2623 | 15-40 | 14 | 1500 |
| 7 | 2012/06/14 12:35 | M1.9 | S17W06 | S26E02 | +0.9 | S27E02 | 987 | 1626 | 15-40 | 14 | ---- |
| 8 | 2012/07/17 12:50 | M1.7 | S28W65 | S27W79 | +4.5 | S32W79 | 958 | 1881 | 40-80 | 136 | 10 |
| 9 | 2012/07/23 01:50 | back | S17W141 | N05W135 | +5.1 | N00W135 | 2003 | 2621 | 165-500 | 12 | 5000 |
| 10 | 2012/08/31 19:24 | C8.4 | S19E42 | S06E40 | +7.2 | S13E40 | 1442 | 1601 | 15-40 | 59 | (3500) |
| 11 | 2012/09/27 23:24 | C3.7 | N06W34 | N16W29 | +6.9 | N09W29 | 1319 | 1479 | 80-165 | 28 | HiB |
| 12 | 2013/03/15 05:42 | M1.1 | N11E12 | N10E08 | -7.2 | N17E08 | 1063 | 1602 | 40-80 | 16 | (0.6) |
| 13 | 2013/09/29 21:43 | C1.2 | N23W25 | N23W29 | +6.8 | N16W29 | 1025 | 1864 | 80-165 | 200 | 0.3 |
| 14 | 2013/12/28 17:16 | back | S08W130 | S01W127 | -2.5 | N02W127 | 1133 | 1918 | 80-165 | 30 | 1 |
| 15 | 2014/01/06 08:00 | C2.1[a] | S15W108 | S06W102 | -3.6 | S02W102 | 1293 | 2287 | >700 | 40 | 3 |

[a]The true flare size is unknown because part of the flare was occulted by the limb. [b]Peak >10 MeV flux detected by GOES; [c]Peak >10 MeV flux derived from STEREO/HET data. The numbers in parentheses are from STB; others are from STA